\documentclass[12pt]{article}
\usepackage{amsfonts}
\begin{document}
\def \ep{\epsilon}
\def \om {\omega}
\def \th{\theta}
\def \intR {\int_{-\infty}^{+\infty}}
\def \grad {\nabla}
\def \ov{\over}
\def \q {\quad}
\def \qq {\qquad}
\def \pd {\partial}
\def \bar{\overline}
\def \beq{ \begin{equation} }
\def \eeq{\end{equation}}
\def\~#1{\widetilde #1}

\title{{\bf Chaos in Black Holes Surrounded by Electromagnetic Fields}}

\author{Manuele Santoprete\thanks{Dept. of Mathematics and Statistics, University of Victoria, 
P.O. Box 3045, Victoria B.C., Canada, V8W 3P4. Email: msantopr@math.uvic.ca.  }~ and 
Giampaolo Cicogna\thanks{Dip. di Fisica ``E.Fermi'' and I.N.F.N., Sez. di Pisa, 
Via Buonarroti 2, Ed. B, I-56127, Pisa, Italy. Email: cicogna@df.unipi.it }}

\date{}

\maketitle

\begin{abstract}

\noindent In this paper we study the occurrence of chaos for charged particles moving
around a Schwarzschild  black hole, perturbed by uniform 
electric and magnetic fields. The appearance of chaos is analyzed
resorting to the  Poincar\'e-Melnikov method.
\end{abstract}
~\qq\qq


\vfill\eject
~
\vskip 2truecm

\baselineskip .56cm
\small\normalsize
\section{ Introduction}

In the last  decade chaotic behaviour in general relativity started to
be the subject of many interesting papers. Two main lines of research can
be recognized. The first deals with chaoticity associated with inhomogeneous
cosmological models, the second line assumes a given metric and looks for
chaotic behaviour of geodesic motion in this background. An interesting
selection of references can be found in \cite{Letelier}.
In particular many papers devoted to the study of chaotic dynamics in
general relativity resort to the Poincar\'e-Melnikov method (see
e.g.~\cite{Melnikov,Wiggins} for the general theory). The Melnikov method is
an analytical criterion to determine the occurrence of chaos in integrable
systems in which homoclinic (or heteroclinic) manifolds biasymptotic
to unstable critical points or to periodic orbits
(more generally to invariant tori) are subjected to small perturbations.
Such perturbations may lead to the phenomenon of transversal intersections
of the stable and unstable manifolds. This kind of dynamics can be then
detected by the Melnikov functions, since they  describe the transversal
distance between the stable and the unstable manifolds of the critical
point or periodic orbit.

The Melnikov method has been applied in many branches of physics
and applied mathematics, so that it is impossible to give here even a
partial account of the vast literature, and we will quote only some of the
applications to general relativity that are more strictly related to the
present paper (for a more complete list of references see
e.g.~\cite{Wiggins,Letelier99,Santoprete}).
Examples of applications of the Melnikov method in general
relativity concern the study of the orbits around a black hole perturbed
either by an external quadrupolar shell \cite{Letelier99, Moeckel} or
gravitational radiation \cite{Letelier,Bombelli}.

In this work we  take a slightly different approach, compared
with the previous literature. Firstly, we  do not analyze perturbations
of the metric, but perturbations due to the interactions
produced by uniform  electric or  magnetic fields.  We  consider these as
perturbations to the Hamiltonian of a charged test particle in
free fall in a Schwarzschild black hole. Secondly, we do not restrict
ourselves to perturbations lying in the plane of the orbit, but
we deal more in general with the full three dimensional problem, i.e. with
perturbations which may change the plane of the orbit.
Let us also remark that we will consider time-independent perturbations,
and show that they produce chaos; this happens even when only the
rotational invariance is broken.

The problem we study in this paper can be related to the
astrophysical reality, since it is well known that magnetic fields can be
associated with black holes.
We will consider Schwarzschild black holes, i.e. non rotating ones, and we
reserve to tackle the more interesting, but more complex problem of
rotating black holes in a forthcoming paper.
Isolated black holes cannot possess  any properties other than mass,
electric charge and angular momentum, but  the medium surrounding the hole
can be responsible for the magnetic field. For example,  super massive
black holes can acquire surrounding matter either by gravitationally
pulling interstellar gas into its vicinity, or by the disruption of passing
stars. The surrounded matter will be shaped in an accretion disk
in a state of plasma, that produces a magnetic field.

Electric fields are less likely to be found near a black hole,
although the Blandford-Znajek mechanism \cite{Blandford} is a process that
can develop a potential difference (i.e. an electric field) between the
poles and the equator of a black hole spinning in a magnetic field
pointing along the axis of rotation.

On the other hand, it is not easy to account for the presence of
charged particles since  the ionized gas, that is shaped in an accretion
disk surrounding a black hole, is in a state of plasma, and a plasma is
electrically neutral. One way to account for the presence of charged
particles is to consider again the Blandford-Znajek mechanism, that, if the
field strength is large enough,  can separate charges and accelerate them
to relativistic velocities. Otherwise, we can model the motion of
electromagnetic currents in a
macroscopic piece of plasma considering the equivalent problem of the
motion of a charged particle moving around a black hole surrounded by a
magnetic field. This can be done since the magnetic field has the same
effect on an electric current in  a macroscopic piece of plasma as on a
single charge.

As said before, we will deal in this paper only with uniform
fields. Although the fields which can be found in the astrophysical
reality, e.g. the Blanford-Znajek ones, are not uniform, some authors
(see \cite{Dokuchaev} and references therein) pointed out  that  models with
uniform external fields are a fairly good approximation in order to explain
the qualitative features of a black hole in a magnetic field, indeed
qualitative arguments indicate the existence of a quasi-uniform field.

In the next section we  discuss the equation of motion for the
Schwarzschild solution, we find the homoclinic orbit and  show that the
perturbed systems we are considering are of type III, according
to the classification given in \cite{Wiggins}. In section 3 we present a
summary of the Melnikov method for a system of this type. In section 4 we
consider the perturbations given by  uniform electric and
magnetic fields. In the last section we apply the Melnikov method and we
prove the occurrence of chaos in the perturbed system  both in the case
of electric and of magnetic field, but with a great difference: the
occurrence of chaos is a first-order effect (in the strength of the
field) in the case of the electric field, whereas it is a second-order
effect in the case of the magnetic field. In particular, the Hamiltonian
perturbed by a constant magnetic field turns out to be integrable at the
first-order, and the first non-vanishing contribution to the Melnikov
integral comes from the second-order term. On the other hand, the
occurrence of chaos in this situation has been already proven
\cite{Karas} by means of numerical arguments based on the study of the
trajectories in the Ernst metric; it can be interesting to point out that
our approach confirms this result using a completely different (analytic)
approach. As another interesting result, we  obtain that in both cases
(electric and magnetic fields) only the components of the fields
on the plane of motion are responsible for the chaotic behavior,
whereas the components normal to the plane are not.

It can be significant to remark that, although the occurrence of chaos
can often be expected since the equations of motion are integrable only
in very special cases, the consequences of chaos may be greatly
relevant and sometimes far-reaching. Indeed, we can observe that chaotic
systems may exhibit a very rich dynamics,
including also regions of stability, periodic orbits,
regions of ergodicity and so on, that can also lead to macroscopically
observable phenomena. We recall, just to mention few examples, that
chaotic defocusing of light might make black holes bright \cite{Levin},
(and hence observable) and that the depletion of the outer asteroidal belt
in the Solar System might be explained with the existence of a chaotic
region \cite{Milani}. Moreover, this last  example also reveals the richness
of chaotic dynamic since  it was shown that the only few asteroids found in
the outer belt are resonant and dynamically protected \cite{Milani1}.

\bigskip
\section{Equations of Motion}

In this section we study the motion of a relativistic (charged) particle in
free fall in a gravitational field.
To describe the motion of a  particle  moving in a  space-time with a
metric $g_{ab}$ it is convenient to consider  the action (with $c=1$):
\begin{equation}  S={m \over 2}  \int ds \qq{\rm  or}  \qq S= {m \over 2}
\int g_{ab}\dot x^a \dot x^b ds \end{equation}
where $x^a(s)$ denotes the world-line of a particle and $m$ denotes the
mass.
Using the same notations as in  \cite{Letelier}, the canonical conjugate
momentum to $x^a$ is $p_a=m g_{ab}\dot x^b$ and satisfies the mass-shell
constraint $g^{ab}p_a p_b=-m^2$. The Hamiltonian of the system can be
defined as:
\begin{equation}  H_0={1 \over 2m}g^{ab}p_a p_b
\end{equation}
Let  our background metric be the Schwarzschild metric, i.e. the
metric of a non-rotating black hole:
\begin{equation} ds^2= -f dt^2+f^{-1} dr^2 + r^2(d\theta^2 +\sin ^2\theta
d\phi^2),
\end{equation}
where  $f= 1-2M/r$. Then the Hamiltonian (2) becomes:
\beq
  H_0= {1 \over 2m}\left( p_r^2 f+{p_\theta^2 \over r^2}+{p_\phi^2
  \over r^2 \sin^2 \theta} \right)
-f^{-1}\left({E\over m}\right)^2
\eeq
where
\begin{equation} p_r=m f^{-1} {dr \over ds} \qq, \qq p_\theta=m r^2 {d\theta
\over ds} \end{equation}
and
\beq
E=-p_t= m f {dt \over ds } \q , \q  p_\phi= mr^2 \sin ^2 \theta {d
\phi \over ds} \q {\rm and} \q L^2=p_\theta^2 +{p_\phi^2 \over
\sin^2\theta}
\eeq
are conserved quantities. We can introduce the following  Hamiltonian,
more suitable for our purposes (i.e. for finding the
orbits of a particle in free fall in the Schwarzschild metric):
\beq
{\cal H}_0=   {E^2-m^2 \over 2m}= {1 \over 2m} \left ( p_r^2
+{p_\theta^2 \over r^2 }+{p_\phi^2 \over r^2 \sin ^2\theta } \right) -
{Mm\over r}- {Mp_\theta^2 \over m r^3}- {Mp_\phi^2 \over m  r^3 \sin^2
\theta},
\eeq
which can also be written in terms of ${\cal H}_0$ as:
\beq {\cal H}_0=  H_0 f+ {E^2 \over 2m}- {mM \over r }
\eeq
This new Hamiltonian defines  a problem similar to the one
discussed in   \cite{Letelier}, but in our case  we study the motion of a
particle  in the three dimensional space. Exactly as in \cite{Letelier},
we can show that our Hamiltonian admits an unstable circular orbit
$\gamma$ in the plane $\theta= \pi /2$, together with an homoclinic loop
(or, to be more precise, a one parameter family of homoclinic loops)
biasymptotic to this orbit; the spherical symmetry of the problem
implies that the same happens for every plane for the origin.

Choosing for simplicity the plane $\theta= \pi/2$, it can be easily shown
that this unstable circular orbit $\gamma$ has radius
\beq r_u= { 6M \over {1+\beta}}, \eeq
where $\beta=\sqrt {1-12M^2m^2/p_\phi^2}<1$, and that there are homoclinic
orbits to this invariant set for $0 \leq \beta <1/2$.
The equations of motion for the homoclinic orbits are:
\beq {dr \over ds}= \pm \sqrt{{2 \over m} \left( {E^2-m^2 \over
2m}-{p_\phi^2 \over r^2  }  + {Mm\over r}+ {Mp_\phi^2 \over m  r^3 }
\right)} \q , \q {d \phi \over ds}= {p_\phi^2 \over m r^2}, \eeq
where the sign $-$ (resp. $+$)  holds for $s<0$ (resp. $s>0$).
Let us denote by
\beq R=R(s) \q {\rm and } \q \Phi=\Phi(s) \eeq
the expressions of $r$ and $\phi$ with respect to $s$ for the homoclinic
orbit with $R(0)=r_{max}$ (the turning point) and $\Phi(0)=\pi$ (i.e.
the orbit having its axis coinciding with the $x$ axis and the point
$r_u$ in the positive $x$).  We can observe  that  the
maximum value of $r$, along this  homoclinic solution, is
$r_{max}=6M/(1-2\beta)$, but  we will not need the explicit expression for
those functions, that  can be found in \cite{Letelier} (at least for
$R=R(s))$,  we only retain, as in \cite{Cicogna,Diacu},  the
information that $R(s)$ is an even function and $\Phi(s)$ is an odd
function of $s$.

In the following sections, we will  study  time-independent
perturbations that destroy the spherical symmetry of the unperturbed
system. This situation is another example, in a completely different
setting, of symmetry-breaking perturbations that were already examined in
\cite{Cicogna,Diacu}.

The perturbations that we  consider in this work can be written
in Hamiltonian form as
\beq  H= H_0 +\ep~ W(r, \phi,\theta, p_r, p_\phi, p_\theta) \eeq
We  want to show that the Hamiltonian in (12) defines  a system of type
III (according to the classification given in
\cite{Wiggins}) with $n=1$ and $m=2$, so that we can apply the Melnikov
theory developed for
this kind of problems to prove that chaos occurs in the perturbed problem.

According to \cite{Wiggins}, systems of type III can be written in the
following general form, where we have denoted by $\om$ (to avoid any
confusion with the azimuthal variable $\phi$) the angle (cyclic) variables
conjugated to the action variables $I$:
\beq
\cases{ \dot x =JD_x H_0(x,I)+\ep
JD_x W(x,I,\om;\ep)\medskip\cr
               \dot I=-\ep D_\om  W(x,I,\om;\ep) \medskip \cr
                \dot\om=D_I H_0(x,I)+\ep D_I  W(x,I,\om;\ep)}
\q
   (x,I,\om) \in \mathbb{R}^{2n}\times \mathbb{R}^{m} \times T^m
\eeq
where $0< \ep \ll 1$, the dot indicates differentiation with respect
to~ $s$, and
$J$ is the standard symplectic matrix  which in our case ($n=1$) is simply
\beq J= \left(\begin{array}{cc} 0 & 1 \\ -1 & 0 \end{array}\right)
\eeq
and where the subsystem, for $\ep=0$,
\beq \dot x=JD_x  H_0(x,I) \eeq
is  assumed to be completely integrable, and to admit a hyperbolic fixed
point with a homoclinic loop connecting this point to itself.

In our case we can write the equations of motion as
$$\displaystyle{
\cases{  \dot r={\partial  H_0 \over \partial p_r}+\ep {\partial  W
\over \partial p_r} \medskip\cr
\dot p _r=-{\partial H_0 \over \partial r} -\ep {\partial  W \over \partial
r}\medskip}}\eqno(16a)$$
and
$$\displaystyle{
\cases{
\dot p_\phi = -\ep {\partial W\over{\partial \phi}}\medskip\cr
\dot p_\theta =-{\partial  H_0\over \partial\theta}-\ep{\partial  W\over
\partial\theta}\medskip\cr
\dot\phi= {\partial  H_0 \over \partial p_\phi} + \ep {\partial  W \over
\partial p_\phi}\medskip\cr
\dot\theta ={\partial  H_0 \over \partial p_\theta}+\ep {\partial  W \over
\partial p_\theta}}}\eqno(16b)
$$
where, again, the dot indicates differentiation with respect to~ $s$.
The subsystem (16$a$) is, for $\ep=0$, precisely as in (15) and in
particular it admits an hyperbolic fixed point, given in (9).
The subsystem (16$b$), here written in terms of the spherical variables
$\phi,\theta$, should be transformed as in the second and third lines of
(13) by introducing two action variables $I$ (e.g., $I_1=p_\phi,\
I_2=L$, the total angular momentum) with their conjugated cyclic
variables $\om_1,\om_2$; we shall see that it is not necessary
to explicitly perform this transformation.

\addtocounter{equation}{1}
\bigskip

\section{The Melnikov Method}
According to \cite{Wiggins}, the general expression of the Melnikov functions
for  a system of type III reduces in our case to
\beq
M_i^{\overline{I}}(\om_0)= -\intR D_{
\om_i} W(q_0^{\overline{I}}(s);0)~ds \qq\q (i=1,2)
\eeq
where
\beq
q_0^{\overline{I}}(s)\equiv\big(x^{\overline I}(s),
\overline{I},\int^t D_I  H_0(x^{\overline{I}}(\xi),\overline{I})d\xi
+\om_0\big)
\eeq
is a generic homoclinic orbit of the unperturbed problem.
$\overline{I}$ has to be chosen so that it defines a KAM
torus (see \cite{Wiggins} for more details). Actually, the convergence
of (17) is a delicate matter: indeed the integrals  converge
only ``conditionally'', i.e. when the limits of integration are
allowed to approach $+\infty$ and $- \infty $ along suitable sequences
${T_j^s}$ and ${-T_j^u}$ respectively. Such sequences must be chosen as in
\cite{Wiggins}, i.e. for every $\epsilon$ sufficiently small we have to
consider  monotonely increasing sequences of real numbers, with
$j=1,2,\dots$ and  $\lim_{j \to \infty} T_j^{s,u}=\infty$, such that
$\lim_{j \to \infty}| q_\epsilon ^s(T_j^s)-q_\epsilon^u(-T_j^u)|=0$ and
$\lim_{j \to \infty} |D_\om (q_\epsilon ^s(T_j^s),0)|=\lim_{j \to
\infty}|D_\om(q_\epsilon^u(-T_j^u),0)|=0$,
where $q_\ep^{s,u}$ are trajectories of the perturbed system
($\ep\not=0$) in the stable and unstable manifolds (see \cite{Wiggins}).

Moreover we recall that the existence of simultaneous zeroes of the
Melnikov functions in (17) is sufficient to prove the occurrence of chaos.
In fact it is known that the presence of simultaneous zeroes (and the
periodicity of the perturbation) implies an infinite sequence of
transversal intersections of the stable and unstable manifold leading to
a chaotic dynamics (see e.g. \cite{Wiggins}).

With  $I_1=p_\phi,\ I_2=L$, and
dropping the superscript over $\overline{I}$ to
simplify the notation,  we then find from (17) the two conditions
\beq M_{p_\phi}=\intR \{p_\phi, W\}\ ds =\ 0. \eeq
and
\beq
M_{L}= \intR \{L,  W \}\ ds\ =\ 0
\eeq
where both integrals, according to (17-18), are to be evaluated along
a generic homoclinic orbit $q_0^I$ (18).

Apart from a rotation, which transforms the given perturbation $W$ into a
new $\~W$ (see below for more details), we can always choose the
homoclinic orbit in the plane $\theta=\pi/2$, therefore it is easily seen
that (19-20) become simply
\beq
M_{p_\phi}=\intR  {\partial \~  W \over \partial \phi }\ ds\ =\ 0
\label{Mphi}
\eeq
and
\beq
M_L=\intR p_\phi\ {\partial \~  W\over \partial \phi}\ ds\ =\ 0
\eeq
Condition (22) turns out to be the same as (21), and then
we are left with only one condition. Hence, to find transversal
intersections of the stable and unstable manifolds to the periodic orbit,
and therefore a chaotic behavior, we have to find simple  zeroes of the
Melnikov function (21). Let us remark incidentally that the transversality
of these intersections is not strictly necessary, indeed -- to have
chaos -- it would be sufficient that the crossing is ``topological'',
i.e. that there is really a crossing from one side to the other
\cite{Burns}.

It is interesting to remark that we have found that, in our case,
only one Melnikov condition has to be studied. The same result would be
clearly obtained for the  same problem restricted to the plane. This shows
that to study the occurrence of chaos in this kind of systems (i.e. with
conserved Hamiltonian and angular momentum) it is sufficient to consider
the  planar problem.

As already stated, to transform the conditions (19-20) into (21-22), a
rotation is necessary, and in particular one may choose the rotation in
such a way that the generic homoclinic orbit $q_0^{(I)}$ is transformed
precisely into the
homoclinic orbit in the plane $z=0$ with axis coinciding with the $x$ axis,
as chosen in Sect. 2 \big(see (10-11)\big). This rotation is defined by
the following Euler angles (with the conventions and notations as in
\cite{Go}):
\beq -\Omega,\ - i,\ - \om
\eeq
where (with the language of celestial mechanics) $i$ the inclination of the
plane of the orbit, $\om$ the angle of the perihelion with the line of nodes
in the orbital plane, and $\Omega$ is the longitude of the ascending node.
Notice that, in terms of our previous variables, one has $i=\,
$arccos $(p_\phi/L)$ and
\beq \Omega=\om_{01}\ ,\  \om=\om_{02}
\eeq
which then play the role of the arbitrary ``phases'' $\om_0$ in eq. (18).

Denoting by $A$ the matrix of this rotation, the given perturbation
$W({\bf x},{\bf p}_{\bf x})$  $\big({\bf x}=(x,y,z)$, etc.\big) will
assume a new expression $\~W$ obtained replacing ${\bf x}$ with
$A{\bf x}$ and ${\bf p}_{\bf x}$ with $A{\bf p}_{\bf x}$. It is then
easy to verify that the Melnikov condition (21) becomes
\begin{eqnarray}
& & M(\om,\Omega) = \\
& & \intR R(s)\Big(- C_1\big(R(s),\Phi(s),\om,\Omega\big)\, \sin \Phi(s)+
C_2(\ldots)\cos\Phi(s)\Big)\ ds = 0 \nonumber
\end{eqnarray}
with
\begin{eqnarray}
C_1&=&\Big({\pd \~W\over{\pd x_1}}\Big)_0(\cos\om\cos\Omega-
\cos i \sin\om\sin\Omega)+  \\
&+&\Big({\pd \~W\over{\pd x_2}}\Big)_0
(-\cos\om\sin\Omega-\cos i\sin\om\cos\Omega)+
\Big({\pd \~W\over{\pd x_3}}\Big)_0\sin i\sin\om \nonumber \\
C_2&=&\Big({\pd \~W\over{\pd x_1}}\Big)_0(\sin\om\cos\Omega+
\cos i \cos\om\sin\Omega)+ \nonumber \\
&+&\Big({\pd \~W\over{\pd x_2}}\Big)_0
(-\sin\om\sin\Omega+\cos i\cos\om\cos\Omega)-
\Big({\pd \~W\over{\pd x_3}}\Big)_0\sin i\cos\om \nonumber
\end{eqnarray}
and where $(\pd \~W/\pd x_i)_0$ means that in the derivative of the given
$W$ with respect to $x_i$ one has to replace ${\bf x}$ with $A{\bf x}$ and
finally put $z=0$ (or $\theta=\pi/2)$.

If, e.g., $i=0$, i.e. if the
problem is completely planar, including the perturbation, or if the
perturbation is ``generic'', i.e. has no ``preferred'' direction in the
space (as often happens, see next Section for explicit examples),
and therefore it is not restrictive to assume $i=0$, then one gets
\beq
C_1=\Big({\pd \~W\ov \pd x_1}\Big)_0\cos\phi_0 +
\Big({\pd \~W\ov \pd x_2}\Big)_0\sin\phi_0\ , \q
C_2=\Big({\pd \~W\ov \pd x_1}\Big)_0\sin\phi_0-
\Big({\pd \~W\ov \pd x_2}\Big)_0\cos\phi_0
\eeq
where $\phi_0=-(\om+\Omega$), the needed rotation is simply a
rotation of angle $\phi_0$  around the $z-$axis and (25) becomes
\beq
M(\phi_0)=\intR{\pd W\ov {\pd
\phi}}(R(s),\Phi(s)+\phi_0,\pi/2,\dot R(s),L,0) \, ds=0
\label{M0}
\eeq
In conclusion, it is clear that verifying Melnikov conditions for the
appearance of
chaotic behaviour amounts to verifying the existence of values of
$\om,\Omega$ for which (25) (or (28) in the above hypothesis) is satisfied.

\bigskip

\section{The Perturbations}
In this section we want to consider the motion of a relativistic charged
particle in a gravitational field with an electromagnetic perturbation.
To this end, we need the Hamiltonian describing a particle in an
electromagnetic field, i.e.
\beq H= {1 \over 2m}g^{ab}(p_a-eA_a)(p_b-eA_b)
\label{Hel}
\eeq
and hence, for weak fields:
\beq
H= H_0-{e \over m} g^{ab} p_a A_b~ +~ {\rm higher\ or der\ terms}
\eeq
Let us remark that we neglect the effect of the electromagnetic
fields back on the metric since direct effects of the magnetic field on a charge are
generally very large compared to indirect gravitational effects on the
mass arising from gravity of the field energy; we recall, incidentally, that the metric of a black
hole immersed in a uniform  magnetic field was found as an exact solution of the
Einstein-Maxwell equations \cite{Ernst, Karas}.

Let us first consider  a  uniform electric field in a generic
direction ${\bf l}$; then
$A_i=(-\psi,{\bf  A})=(-{\cal E} \ \hat n \cdot {\bf l} ,0)=(-{\cal E}(l_1
\cos\phi\sin\theta+ l_2 \sin\phi \sin\theta +l_3 \cos\theta),0)$,
where $\hat n $ is the unit vector in the ${\bf r}$ direction.  The
perturbed Hamiltonian (30) is then
\beq H=H_0-\ep Ef^{-1}  (l_1 \cos\phi\sin\theta+ l_2
\sin\phi \sin\theta +l_3 \cos\theta) \eeq
where $\ep= e{\cal E}/m \ll 1  $ .

More interesting, and a little more complicated, is the case
where the perturbation is given by a magnetic field. Let $A_i=(0,{\bf
A})=(0,{1\over 2} {\bf B} \times {\bf r})$ where:
\beq
{\bf A}= {B \over 2}~(k_2 z-k_3 y, k_3 x -k_1 z, k_1 y-k_2 x),
\eeq
with ${\bf B}= B~{\bf k} =B~(k_1, k_2, k_3)$, where ${\bf k}$ is a unit
vector in the direction of the magnetic field. Rewriting  the vector
potential in spherical coordinates we obtain
\beq
\left \{
\begin{array}{l}
A_r=0 \\
A_\phi={ Br^2 \over2}(k_3\sin^2\theta-\sin\theta \cos\theta(k_1 \cos\phi+
k_2 \sin\phi)).\medskip \\
A_\theta= {Br^2 \over 2}(k_2 \cos\phi -k_1 \sin\phi)  \\
\end{array}
\right.
\eeq
In both cases the direction of the perturbing field is generic, therefore
no rotation is required (it would simply change the directions ${\bf
l},\ {\bf k}$, which are not fixed; see the remark at the end of previous
section).

The perturbed Hamiltonian (29) can be written as
\begin{eqnarray}
H=H_0 - {B e r^2 \over 2m }
\left[ g^{\phi\phi}p_\phi(k_3\sin^2\theta-
\sin\theta \cos\theta(k_1 \cos\phi+ k_2 \sin\phi))\right]- \nonumber \\
-{B e r^2 \over 2m}\left[ (g^{\theta \theta}p_\theta
(k_2 \cos\phi-k_1\sin\phi)) \right]+ \nonumber\\
+ {B^2e^2r^4 \over 8m} \left[ g^{\phi\phi}(k_3\sin^2\theta-
\sin\theta \cos\theta(k_1 \cos\phi+ k_2 \sin\phi))\right]^2 +\nonumber\\
+ {B^2e^2r^4 \over 8m} \left[ (g^{\theta \theta}(k_2 \cos\phi-k_1\sin\phi))
\right]^2
\end{eqnarray}
where this time we retain also the quadratic terms in the magnetic field, in
view of the discussion in the next section.  Since we need in (\ref{Mphi})
(or (\ref{M0})) quantities evaluated along the homoclinic
orbit in the plane $\theta=\pi/2$, we can rewrite, recalling that
$g^{\phi\phi}=r^{-2}(\sin\theta)^{-2}$ and $g^{\theta\theta}=r^{-2} $,
the Hamiltonian (34) as:
\beq
  H= H_0-\epsilon  k_3~p_\phi + \ep^2 {mr^2 \over 2}
  \left[(k_1^2-k_2^2)\sin 2\phi-k_1k_2 \cos 2\phi \right ]
\eeq
where $\epsilon={Be \over 2m} \ll 1$.

\bigskip

\section{The Melnikov Conditions}

The Melnikov integral for the perturbation produced by the uniform
 electric field  can be found from (\ref{Mphi}) (or (\ref{M0})):
\beq M_\phi =\intR Ef^{-1}(l_1
\sin(\Phi(s)+\phi_0)-l_2\cos(\Phi(s)+\phi_0)) \ ds \eeq
since $\theta= \pi/2$. Now, using the fact that $R$
and $\Phi$ are respectively even and odd functions of  $s$,  we can
write the integral as:
\beq
  M_\phi=(l_1\sin\phi_0 -l_2\cos\phi_0) \intR Ef^{-1} \cos\Phi(s)~ ds.
\label{Mphi1}
\eeq
or, defining two constants $L$ and $
\alpha $ such that $l_1=L \cos\alpha $ and $l_2=L\sin\alpha $, the
Melnikov condition is
\beq
M_\phi=J_1 L \sin(\phi_0-\alpha)=0
\eeq
where
\beq
J_1=J_1(p_\phi, M,m)= \intR Ef^{-1} \cos\Phi(s)~ ds
\label{MJ1}
\eeq
 From numerical evaluations and general arguments for this type of
(conditionally convergent) integrals (see also \cite{Letelier}), we can
assume that $J_1 \neq 0$, or that it vanishes for at most some isolated
values of the parameters ($p_\phi,m,M$) involved. Therefore,  the
Melnikov function (\ref{Mphi1}) has simple zeroes and hence, thanks to the
periodicity of the perturbating term in the integral, there is a infinite
sequence of transversal intersections of the asymptotic stable and unstable
manifolds, leading as well known to a chaotic dynamics.

In the case of the magnetic field, it is clear from (35) and (28) that
the first-order term in $\ep$ of the perturbation gives no contribution
to the Melnikov integral: actually, it is easy to see that the perturbed
Hamiltonian truncated at the first-order is integrable (indeed, it admits
$L^2$ as an additional constant of motion). Then, if we consider the
first non-vanishing contribution to the Melnikov integral, this gives the
following condition
\beq
  [(k_1^2 - k_2^2) \sin 2\phi_0 - 2 k_1k_2 \cos 2\phi_0]
     \intR R^2(s) \cos 2\Phi(s) ds =0
\label{Mphi2}
\eeq
which can also be written, with obvious notations, similar to the above
(37-39),
\beq
J_2K\sin(2\phi_0-\delta)=0
\eeq
As before, we can conclude that chaos occurs also in this case, but --
unlike the case of electric field -- this is now a ``second-order effect''.
On the other hand, the  motion of test particles moving around a black hole
immersed in a magnetic field was studied in \cite{Karas}, where the
authors, performing a  numerical study
of  the orbits of a particle in  Ernst space-time, presented  strong evidence
of the occurrence of chaos and nonintegrability. As the authors acknowledge,
the  numerical methods they use cannot give a
rigorous proof of the nonintegrability or of the occurrence of chaos.
Instead, the Melnikov method is able to detect the existence of
a infinite sequence of transversal intersections
of the stable and unstable manifolds and hence, via the Smale-Birkoff theorem,
to manifest the equivalence to a symbolic dynamics expressed by the Smale
horseshoe. Therefore, the Melnikov technique provides an analytic proof of
the occurrence of chaos in the problem discussed here and in \cite{Karas}.
In particular our analysis shows that the chaotic dynamics appears even
when the reaction of
the magnetic field on the black hole is neglected, but does not when
the terms quadratic in the magnetic field  (\ref{Hel}) are neglected.

Finally, it can be observed that, in both cases (i.e. for both the electric
and magnetic field) whereas the component of  the  field  on the plane
of motion  leads to a  chaotic
dynamics, the component normal to the plane does not. This behavior can be
explained observing  that the component  normal to the plane
of motion is a constant on such plane. Therefore it  doesn't break the
symmetry of the system on the plane of motion,
and hence it doesn't lead to the appearance of chaos.
Since the problem is spherically symmetric,  the
reasoning used for the $\theta=\pi /2$ plane can be applied to every plane
for the origin of coordinates. Thus,  given an electric or magnetic
field,  on each  plane for the origin (except at most  the one normal to
the field),  chaos appears for a suitable choice of initial
conditions.

\qq\qq

\noindent
{\bf Acknowledgments}  The first author (MS) is grateful to Professor Werner
Israel for his enlightening  comments and suggestions,
and to the University of Victoria for financial support (University of
Victoria Fellowship).

\vfill\eject



\begin{thebibliography}{2001}


\bibitem{Letelier}
Letelier P.S. and Vieira W.M. (1997)
{\it Class. Quantum Gravity} {\bf 14}, 1249 


\bibitem{Melnikov}
Melnikov V.K.  (1963) {\it Trans. Moscow Math. Soc.}, {\bf 12}, 1 

\bibitem{Wiggins}
Wiggins S. (1988) {\it Global Bifurcations and Chaos} (New York, Springer)

\bibitem{Letelier99}
Letelier P.S. and Motter A.E. (1999) {\it Phys. Rev.  E}, {\bf 60}, 3920 


\bibitem{Santoprete}
Santoprete M. (1999) { \it Thesis}, Dept. of Physics, Univ. of Pisa


\bibitem{Moeckel}
Moeckel R. (1992) { \it Comm. Math. Phys.}, {\bf 15}, 415

\bibitem{Bombelli}
Bombelli L. and  Calzetta E. (1992) {\it Class. Quantum Gravity}, {\bf 9}, 2573

\bibitem{Blandford}
Blandford R.D. and Znajek R.L. (1977) {\it Mon. Not. R. Astr. Soc.},
{\bf 179}, 433

\bibitem{Dokuchaev}
Dokuchaev V.I. (1987) {\it Sov. Phys. JETP}, {\bf 65}, 1079

\bibitem{Karas}
Karas V. and Vokrouhlick\'y D. (1992) {\it Gen. Rel. Grav.}, {\bf 24}, 729

\bibitem{Levin}
Levin J. (1999) {\it Phys. Rev. D}, {\bf 60}, 064015

\bibitem{Milani}
Milani A. and Nobili A.M. (1985), {\it Astron. Astrophys.}, {\bf 144}, 261

\bibitem{Milani1}
Milani A. and Nobili A. M. (1984), {\it Cel. Mech.}, {\bf 34}, 343

\bibitem{Cicogna}
Cicogna G. and Santoprete M. (2000) {\it J. Math. Phys.}, {\bf 41}, 805 

\bibitem{Diacu}
Diacu F. and Santoprete M. (2001) {\it  Physica D}, {\bf 156}, 39 

\bibitem{Burns}
Burns K. and  Weiss H. (1995) {\it Comm. Math. Phys.}, {\bf 172}, 95 

\bibitem{Go}
Goldstein H.  (1980) {\it Classical Mechanics} (Reading, Ma,
Addison-Wesley)

\bibitem{Ernst}
Ernst, F. J. (1976), {\it J. Math. Phys.}, {\bf 17}, 54


\end{thebibliography}
\end{document}